\documentclass{article}

\usepackage{arxiv}

\usepackage[utf8]{inputenc} 
\usepackage[T1]{fontenc}    
\usepackage{hyperref}       
\usepackage{url}            
\usepackage{booktabs}       
\usepackage{amsfonts}       
\usepackage{nicefrac}       
\usepackage{microtype}      
\usepackage{lipsum}
\usepackage{graphicx}
\graphicspath{ {./images/} }

\title{Transportation Transformed: A Comprehensive Review of Dynamic Rerouting in Multimodal Networks}

\author{
 Suyash Pratap \\
 Department of Industrial Engineering \\
 Kansas State University \\
 2061 Rathbone Hall, 1701B Platt St, Manhattan, KS 66506 \\
  \texttt{spratap@ksu.edu} \\
}

\begin{document}
\maketitle
\begin{abstract}
The emergence of dynamic rerouting in multi-modal transportation networks has emerged as a crucial area in operations research, revolutionizing routine optimization. The review study analyzes multiple research publications on algorithms and techniques related to dynamic rerouting to give a thorough summary of the state of research in this field and provide future suggestions. The research paper explains the importance of dynamic rerouting in modern transportation systems and recognizes its critical role in tackling issues like accidents, traffic congestion, and infrastructure constraints. In addition, the review examines the development of dynamic rerouting techniques by examining several studies to uncover the theoretical foundation, technological developments, and effects of the practices on various forms of transportation. The paper emphasizes the potential of technological advancements such as artificial intelligence, the Internet of Things, and big data in transforming routing efficiencies. Further, the review presents specific difficulties and best practices for each mode of transportation, highlighting the many uses of dynamic rerouting in air, sea, rail, and road transportation. The review also digs deeper into the integration barriers common in multi-modal networks, highlighting successful case studies that overcome these obstacles as well as strategic approaches and regulatory modifications. Lastly, the research paper assesses the impact of dynamic rerouting on urban development, sustainability, and potential directions for future research such as the integration of large language models. The comprehensive literature review incorporates multiple research perspectives to offer significant insights into the efficacy, challenges, and potential future pathways for dynamic rerouting within multi-modal transportation networks. 
\end{abstract}

\section{Introduction}
In modern transportation, multimodal networks enable the movement of people and products across roadways, railways, air routes, and sea channels. Global trade and social connectivity depend on these networks, which seamlessly integrate modes for efficient and complete transit solutions. Notably, dynamic rerouting is vital in optimizing multimodal transportation networks \cite{zukhruf2022developing}. The concept of dynamic rerouting marks a significant shift as it mirrors current conditions and unforeseen occurrences, allowing for prompt alterations to transit routes and modes in response to changing situations. In addition, dynamic rerouting maximizes resource efficiency, decreases transit times, and improves system resilience by addressing unexpected disturbances, including traffic congestion, accidents, weather, and infrastructure maintenance \cite{darvishan2021dynamic}. Flexibility in transportation infrastructure using real-time data, predictive analytics, and adaptive algorithms is another key advantage of dynamic rerouting. The evolution of route planning in transportation networks has seen remarkable advancements, making it possible to compute driving directions in milliseconds, even on a continental scale. This progress is attributed to a variety of techniques that balance preprocessing effort, space requirements, and query time \cite{bast2016route}. Modern transportation management relies on dynamic rerouting, which combines technology innovation, operational efficiency, and quick decision-making. Therefore, the review paper examines dynamic rerouting in multimodal transportation networks to determine its ramifications and growing importance in transportation logistics and operations.

The increased complexity and demands of transportation make the analysis of dynamic rerouting in multimodal networks urgent and relevant. Globally, multimodal transit systems are becoming more popular, requiring flexible, sensitive, and efficient routing algorithms. Statistics demonstrate the need for dynamic rerouting. For instance, delays caused by congestion cost large economies billions of dollars annually, highlighting the economic consequences of poor routing \cite{muneera2018economic}. In addition, real-time rerouting is crucial for supply chains and passenger convenience due to the increasing frequency of harsh weather and unforeseen delays. Remarkably, in the age of rapid technological growth, big data analytics, AI, and IoT in transportation represent a new frontier in multimodal transportation. Predictive dynamic rerouting allows proactive adjustments to reduce congestion and optimize transit routes, outdoing traditional models. The analysis shows that dynamic rerouting is crucial to solving transportation problems and leveraging new technology. Therefore, the systemic review paper demonstrates the role of dynamic rerouting in multimodal networks, such as increased operational efficiency, resilience, and sustainable transportation systems. 

The review paper comprehensively covers dynamic rerouting in multimodal transportation networks in distinct and clear sections. Understanding dynamic rerouting in transportation is the first section of the research paper, which explains the concept of dynamic rerouting and its relevance in transport. The section explores the evolution practices of dynamic rerouting and discusses the theoretical underpinnings and principles of dynamic rerouting. The paper also contains a section that explores technological innovations in dynamic rerouting. The segment discusses how big data, AI, and IoT improve dynamic rerouting by highlighting the pros and disadvantages. The application of dynamic rerouting across multiple modes of transport is an essential and practical section of the paper. The research examines dynamic rerouting in road, rail, aviation, and maritime modes. The paper also contains a section on algorithms in dynamic rerouting, which focuses on the development of methodology and optimization techniques in different scenarios. Thereafter, the paper discusses issues, solutions, and successful implementations of each mode of transport. Further, the segment of integration Struggles in multimodal networks examines common multimodal network integration issues and how to overcome them. The part also highlights successful case studies to demonstrate how the obstacles were addressed. The section on case studies on dynamic rerouting presents case studies from various areas and transport types to analyze outcomes, lessons gained, and dynamic rerouting solution scalability. The paper provides an impact of dynamic rerouting on urban sustainability and development. The section discusses how dynamic rerouting helps urban transport sustainability, assesses its impact on urban growth, and provides support claims with studies or statistics. Finally, the review paper's last section highlights dynamic rerouting research gaps and future prospects. The last segment discusses dynamic rerouting's future technologies and methods and the identification of research gaps and opportunities for additional study, including the integration of Large Language Models which can enhance multimodal transportation by streamlining communication, improving predictive analytics, and providing advanced support in decision-making and customer service. In conclusion, the section ends with an action plan and suggestions for future research.

\section{Understanding dynamic rerouting in multimodal transportation}
\label{sec:headings}
Dynamic rerouting is crucial to multimodal transportation networks, allowing real-time route and mode changes. Dynamic rerouting involves changing specified paths or modes of transportation in reaction to unexpected events, disturbances, or transit conditions. Notable advantages of dynamic rerouting include minimized disruptions, reduced travel time, efficiency, and ensures smooth movement across transport modes \cite{chan2005behaviors}. Modern transportation paradigms depend on this dynamic flexibility, which provides agility in reaction to changing conditions and optimizes transportation networks. Dynamic rerouting algorithms provide adaptation and efficiency in multimodal transportation networks as they become increasingly complicated and interconnected.

\begin{figure}
    \centering
    \includegraphics[width=15cm]{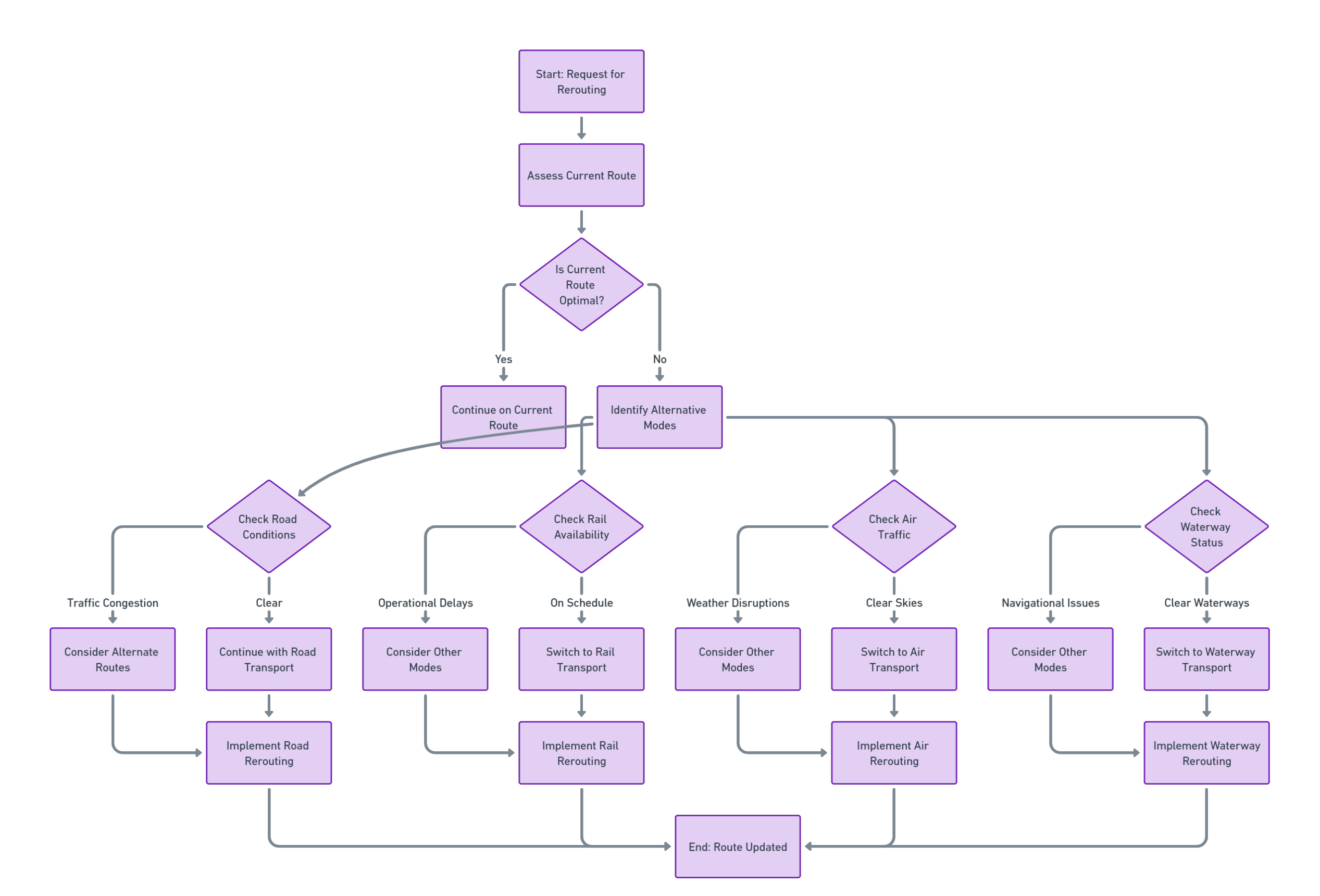} 
    \caption{A generalized multimodal dynamic rerouting system}
    \label{fig:fig1}
\end{figure}

Technological advances and socioeconomic changes have transformed dynamic rerouting methods. Traditional dynamic rerouting relied on manual processes and human interaction to handle disturbances and unprecedented events. Consequently, rerouting was difficult in early transportation systems due to the lack of sophisticated algorithms and real-time data. However, computing advances and data-driven methods changed dynamic rerouting over time. Technological advances created algorithms and computer models that could process massive real-time data. Modern technologies that automatically analyze traffic, weather, and transit data transformed dynamic rerouting with transportation networks, changing from reactive rerouting to proactive systems that could forecast and prevent problems. Technology such as cell phones and GPS systems also accelerated dynamic rerouting \cite{darvishan2021dynamic}. Consequently, users have real-time information to make decisions, and transportation systems can communicate with travellers due to technological developments. In modern times, dynamic rerouting has become a customer-centric solution considering end-user experience and preferences, marking a significant advancement in multimodal transportation network dynamic rerouting.

The theory underpinnings of dynamic rerouting optimize routes in real-time to handle unexpected disruptions and changing conditions in multimodal transportation networks. This requires operations research, algorithm, and network theory ideas. Significantly, optimizing alternate routes or means of transport during disruptions is vital to dynamic rerouting \cite{archetti2022optimization}. Dynamic rerouting systems use these algorithms to recalculate routes to avoid traffic, accidents, and other obstacles quickly. For instance, Dijkstra's or A* search algorithms help determine the shortest paths depending on journey time, congestion, or user preferences \cite{yu2019anti}. Further, network theory is essential to understanding transportation networks and creating effective rerouting solutions. In addition, graph theory provides a solid framework For multimodal networks with nodes representing locations and edges of transportation routes. By understanding network structures and dynamics, dynamic rerouting systems can find alternative routes to avoid disruptions and maximize travel efficiency. Dynamic rerouting systems also use stochastic modeling and predictive analytics to predict disturbances \cite{guo2020dynamic}. Predictive algorithms analyze past data to predict congestion, weather, and infrastructure concerns, allowing proactive rerouting. The theories support dynamic rerouting's goal of optimizing routes in real-time to provide commuters with efficient, dependable, and sustainable travel experiences across multimodal transport networks.
=

\section{Technological Innovations in Dynamic Rerouting}
Modern technology has improved dynamic rerouting in multimodal transportation networks. Advanced technologies improve rerouting speed, precision, and adaptability. Real-time data feeds from sensors, and IoT devices in transportation networks are key breakthrough innovations in dynamic routing \cite{zantalis2019review}. Notable examples of real-time devices include GPS trackers, traffic cameras, weather sensors, and mobile apps that provide constant data. Subsequently, real-time data powers dynamic rerouting systems, which may detect traffic congestion, accidents, and infrastructure faults. Additionally, machine learning and AI algorithms are other essential technological innovations in dynamic rerouting \cite{zantalis2019review}. The systems estimate traffic, find ideal routes, and react to changing conditions using historical and real-time data for efficient transportation. Reinforcement learning algorithms learn from prior events and feedback loops to improve rerouting accuracy and adaptability. Dynamic rerouting has also advanced due to innovative city projects. Traffic management centers and cars communicate via intelligent traffic signals and connected vehicles. The linkage allows dynamic rerouting technologies to send commuters real-time updates and alternate routes, assuring smooth navigation during delays. Further, commuters can now use dynamic rerouting thanks to mobile apps and guidance platforms. The applications offer real-time traffic updates, route ideas, and alternative transportation modes to help users plan their trips using current conditions. Therefore, technological advancements demonstrate dynamic rerouting systems have become adaptable, data-driven, and user-centric solutions to optimize travel experiences in multimodal transportation networks.

In multimodal transportation networks, technological innovations such as big data, AI, and the IoT improve routing efficiency in dynamic rerouting systems. Big data is the collection of numerous sets of data \cite{bormida2021big}. In dynamic rerouting systems, big data refers to the vast and diverse information pool from various sources, such as traffic sensors, mobile devices, weather forecasts, and historical traffic patterns. Subsequently, dynamic rerouting systems make decisions using the depth of information from big data. The vast information helps optimize routes in real-time by revealing traffic flow, congestion hotspots, trip times, and other vital metrics that aid in choosing the most ideal route. Additionally, the use of AI technologies, such as machine learning and predictive analytics, can improve dynamic rerouting techniques \cite{zantalis2019review}. The AI algorithms can predict traffic patterns, anticipate disruptions, and suggest the best alternate routes using large amounts of information. Remarkably, reinforcement learning allows these systems to adapt and improve based on transportation network feedback, improving routing accuracy and adaptability. Further, IoT system, including connected devices and sensors in vehicles, infrastructure, and urban environments, drives dynamic rerouting. Notably, IoT sensors update traffic, weather, and road events in real-time enabling cars, transportation hubs, and traffic management centers to communicate vital rerouting information \cite{zantalis2019review}. Therefore, big data gathers different data, AI processes and predicts patterns and IoT devices deliver real-time updates. The integration of the technology trends allows dynamic rerouting systems to adapt to changing conditions, optimize routes in real-time, and alert commuters, improving routing efficiency in multimodal transportation networks.

The technological advancements such as big data, AI, and the IoT for dynamic rerouting in multimodal transportation networks have advantages and disadvantages. All the technology trends improve decision-making. Big data provides detailed traffic patterns for rerouting methods while AI optimizes routes and predicts disruptions in real-time, and IoT devices provide instant and precise traffic data, which enables proactive decision-making. The three technologies offer adaptability in real-time. Big data allows real-time analysis and speedy adaptability to changing situations, while AI helps systems adapt by learning and improving techniques, and IoT provides real-time updates to help adapt route planning quickly. These technological innovations promote efficiency and optimization \cite{archetti2022optimization}. Big data helps find and use optimal routes using historical and real-time data, while AI optimizes route planning by evaluating multiple elements for more effective travel, and IoT provides instant updates to reduce delays and congestion, improving efficiency. Despite the numerous advantages of Big data, AI, and the IoT, the technological innovations have several limitations. The technology trends face privacy and security issues \cite{alansari2018challenges}. The use of large amounts of personal data in Big data raises privacy concerns, while AI poses security vulnerabilities due to data-driven biases in decision-making, and IoT devices may encounter data breaches to increase security vulnerabilities. Reliance on data accuracy is another limitation of the three technologies. While using big data, data quality and accuracy greatly affect forecast accuracy. In AI, predictive powers depend on training data accuracy and quality, while sensor and device dependability affects accuracy, and erroneous data might mislead routing decisions while using IoT. Lastly, the technologies face infrastructure and cost issues \cite{alansari2018challenges}. Notably, big data uses expensive infrastructure and tools to process data efficiently, while implementing AI systems requires significant computer power and technological infrastructure. Finally, the costly deployment and maintenance of IoT devices across transportation networks is a challenge.

\section{Algorithms in Dynamic Rerouting}
The development of dynamic rerouting algorithms has advanced significantly in transportation and communication networks due to technological advancements. The traditional algorithms such as Dijkstra's and Bellman-Ford laid the groundwork for dynamic rerouting under less complex network conditions. These initial approaches paved the way for advanced, immediate responsive algorithms. While Dijkstra's algorithm is a cornerstone for shortest path problems in graph with non-negative edge weights, Bellman-Ford's algorithm extends the shortest path problem to graphs with negative edge weights. However, in dynamic rerouting scenarios, where network conditions are constantly changing and unpredictable, static algorithms are not sufficient. One commonly encountered dynamic rerouting problem in the transport field is the earliest arrival problem, which involves finding the shortest path between a departure station and an arrival destination, taking into account real-time traffic conditions and potential road closures. 

\begin{figure}
    \centering
    \includegraphics[width=12cm]{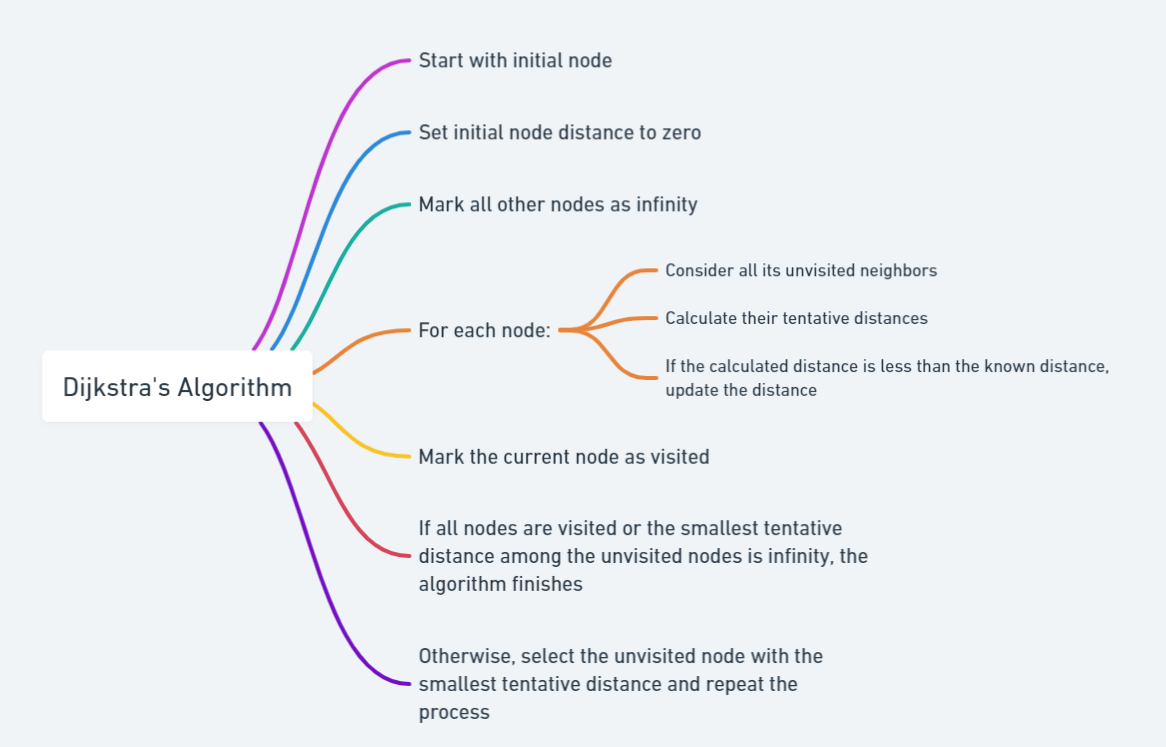} 
    \caption{Dijkstra's Algorithm}
    \label{fig:fig2}
\end{figure}

\begin{figure}
    \centering
    \includegraphics[width=15cm]{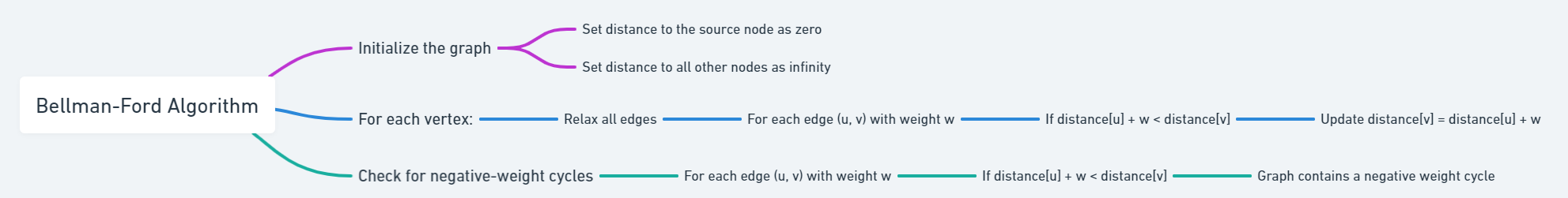} 
    \caption{Bellman Ford's Algorithm}
    \label{fig:fig3}
\end{figure}

Since these fixed algorithms were inadequate to solve many real-world problems, more adaptable rerouting algorithms developed to handle the intricacies of dynamic and changing network conditions. Algorithms like A* were developed which was an extension of the traditional Dijkstra's algorithm by introducing heuristics to guide the search towards the goal node more efficiently. It strikes a balance between the thoroughness of Dijkstra’s approach and the efficiency needed for real-time applications, making it a popular choice in GPS navigation systems \cite{HartPeter}. Yen’s K-Shortest Path Algorithm is another algorithm commonly used in dynamic rerouting scenarios. Developed by Jin Y. Yen in 1971, this algorithm enhances routing path diversity, crucial for network resilience and traffic management. It iteratively finds multiple shortest paths and is particularly useful in scenarios where backup routes are essential \cite{YenKshortest}. Inspired by the foraging behavior of ants, Ant Colony Optimization algorithms have also been applied to dynamic rerouting problems. This algorithm is especially effective in complex and frequently changing environments like urban traffic systems and network routing \cite{DorigoAnt}. Other metaheuristic approaches, such as genetic algorithms and particle swarm optimization, have also shown promise in dynamic rerouting scenarios.These metaheuristic approaches use evolutionary principles and swarm behavior to explore and adapt to the changing network conditions, providing more flexible and responsive solutions.

Recent research has led to innovative algorithms and solutions, particularly in the wake of advanced computational capabilities and data analytics. These algorithms combine real-time data, machine learning techniques, and predictive models to make informed decisions about rerouting, taking into account factors such as traffic congestion, accidents, weather conditions, and road closures. Graph Neural Networks (GNNs) are being utilized to model complex network structures and predict traffic patterns, aiding in more accurate dynamic rerouting decisions \cite{10.1145/3459637.3481916}. Multi-Agent Reinforcement Learning algorithms are applied in dynamic rerouting for coordinating multiple agents (vehicles, traffic lights) in a shared environment, enhancing overall network efficiency \cite{app11114948}. Quantum computing algorithms are also being explored for dynamic rerouting, harnessing the power of quantum mechanics to solve complex optimization problems more efficiently in real-time scenarios. The dynamic system optimal routing model for multimodal transit systems represents a significant advancement, addressing congestion issues in traditional shortest-path-based systems \cite{ma2013dynamic}.

\section{Application of Dynamic Rerouting among Various Modes of Transport}

Dynamic rerouting tactics are used on roads, rail, air, and sea to respond to changing conditions, increase efficiency in operations, and improve transit experiences. In road transport, road networks need dynamic rerouting to manage traffic congestion, accidents, and blockages. Therefore, dynamic rerouting navigation apps provide real-time alternative routes. In addition, through intelligent traffic signals, dynamic rerouting adjusts traffic lights to traffic levels, improving vehicle flow. Further, in rail transport, dynamic rerouting optimizes train timetables for unexpected events like track maintenance or disruptions using railway scheduling and control \cite{rosyida2018literature}. Also, dynamic rerouting allows adaptive signalling systems to modify train routes or speeds for safety or efficiency through adaptive signalling systems in railways. In air transport, dynamic rerouting is used by air traffic control systems to redirect flights during adverse weather and airspace congestion. Most importantly, airlines use dynamic rerouting for flight panning since weather and air traffic congestion affect flight planning. Lastly, vessel traffic services in maritime transport utilize dynamic rerouting to guide ships around congested or dangerous regions, guaranteeing safe journeys. Dynamic rerouting manages vessel arrivals and departures in ports during emergencies or changing situations \cite{rosyida2018literature}. Finally, dynamic rerouting uses real-time data, powerful algorithms, and communication technology to modify route schedules in all modes of transport quickly. The adaptability improves safety, efficiency, delays, and transit reliability.

Dynamic rerouting encounters unique issues across the different transport modes, requiring specific solutions for successful implementation. Road transport faces problems such as unexpected road congestion. The solution to traffic congestion includes real-time traffic monitoring, prediction algorithms, and adaptive traffic signals to rerouting vehicles efficiently \cite{zhang2021review}. In addition, road transport is likely to face limited alternative routes. The solution is finding and using alternatives to avoid congestion and give drivers feasible routes. Rail transport also experiences numerous issues. For instance, railways have permanent rails, making rerouting difficult. However, advanced scheduling algorithms and efficient communication systems allow train schedules or route changes quickly. Transport by rail may face interruptions due to track maintenance. The delays necessitate dynamic rerouting to minimize interruptions \cite{rosyida2018literature}. Regular maintenance planning and predictive analytics reduce these issues. Further, air transport faces numerous challenges. For instance, airlines face flight delays resulting from airspace congestion. The congestion is reduced using collaborative air traffic management technologies and real-time weather and traffic density rerouting. Airlines' flight plans are dependent on weather; thus, flights are negatively affected by adverse weather conditions. However, advanced weather prediction models and alternate flight path planning help dynamic rerouting handle weather disruptions. Further, maritime transport faces challenges such as navigational hazards like storms and shallows, which can block maritime routes. Automated navigation technologies provide real-time data to let ships navigate safer, solving navigational hazards. Lastly, airports experience airport congestion. Dynamic rerouting through efficient vessel scheduling addresses port congestion requirements. Each method of transport requires unique technological advances and adaptive tactics to overcome its challenges. Successful dynamic rerouting implementations across transportation networks require real-time data, predictive algorithms, and efficient communication systems.

Implementations of dynamic rerouting across different modes of transportation modes show its practicality and efficacy. Dynamic rerouting has been successfully implemented in different countries. For instance, the implementation of Expressway Monitoring and Advisory System (EMAS) sensors and cameras in Singapore monitor accidents and congestion \cite{yewemas}. The real-time updates facilitate quick response and information on alternate routes change traffic paths, saving significant travel time. Dynamic rerouting has also been successfully implemented in rail transport. Europe's ERTMS standardizes train control and command \cite{filip2018framework}. It optimizes train routes, improves safety, and boosts rail network efficiency using dynamic rerouting. Further, dynamic rerouting has been successfully implemented in air travel. For example, satellite-based navigation and autonomous dependent surveillance optimize air traffic flow under the US FAA's NextGen system. Dynamic rerouting using weather and air traffic data reduces aircraft delays \cite{sridhar2016towards}. Lastly, shipping giant Maersk optimizes routes with dynamic rerouting solutions based on weather predictions and real-time data to reduce fuel consumption and weather delays, signifying the successful implementation of dynamic rerouting in air travel.

\section{Integration hurdles in multimodal networks}
Different systems, protocols, and logistical issues often cause multimodal network integration issues. However, it's important to consider that the cost factors in multimodal transportation, especially in the context of dynamic alliances, play a critical role in the efficiency and feasibility of these networks. These factors include staff and equipment costs, transportation network considerations, and force majeure factors, all of which collectively determine the cost dynamics of multimodal transport \cite{zhu2019multimodal}. Common challenges of integration hurdles in multimodal networks in technology incompatibility. Transport modalities use different technologies and standards, making communication and coordination between these varied technologies is difficult. In addition, different modes generate and use diverse data sets. Therefore, syncing and integrating data across modes to ensure accuracy, consistency, and real-time availability is difficult, creating data synchronization challenges. Additionally, transportation systems lack interoperability standards, which hinders collaboration. Limited infrastructure is an integration hurdle in multimodal networks since different railway track gauges and road conditions restrict easy transitions and interoperability.  Further, regulatory and policy barriers hinder multimodal network integration since the disparities in policies, rules, and jurisdiction might impede coordination, making integration difficult. Lastly, multimodal timetables, routes, and operations require sophisticated planning and synchronization. A mode change might affect the entire network, creating challenges during operations and coordination. '

Multimodal network integration challenges require a diversified approach to address the shortcomings. Policy changes such as standardizing transport regulations and policies help integration \cite{diao2019towards}. Frameworks that promote interoperability reduce regulatory hurdles. Upgrading infrastructure for seamless mode changes enhances multimodal integration. For instance, standards for train gauges, road signage, and loading facilities can improve intermodal transitions. In addition, technological integration aids in combining multimodal networks. The creation of interoperable products that let transportation systems share data using shared communication platforms improves coordination. To ensure system interoperability and synchronization, industry players should standardize data formats and exchange methods, and centralized databases or platforms for shared data should increase information flow \cite{guo2020dynamic}. Further, training and capacity building facilitate multimodal network integration. Training staff for integrated operations and communicating integration benefits is essential for integrating different networks. The strategy increases public awareness and support for multimodal transportation and integration initiatives aid in enhancing integration of different networks. Therefore, multimodal networks can overcome integration challenges and become more efficient and connected, enhancing mobility and sustainability.

Several case studies demonstrate approaches that countries implement to overcome multimodal transportation integration challenges. First, Singapore's Land Transport Master Plan has integrated many transportation modes through strategic planning \cite{diao2019towards}. The Land Transport Master Plan implements regulations and builds infrastructure to connect buses, railroads, and cycling trails. This integration has significantly decreased congestion and increased efficiency. The Port of Rotterdam, Netherlands, integrates maritime trade with rail and road networks \cite{gurzhiy2021port}. Intermodal terminals and infrastructural modifications by the port authority make shipping, rail, and trucking easier. In addition, Scandinavian Rail Network: Sweden and Norway share a cross-border rail network \cite{marti2013european}. Interoperable rail networks provide smooth passenger and freight mobility, exhibiting international collaboration and standardization. Hamburg's automated traffic control system incorporates vehicles, buses, bicycles, and pedestrians. The technology optimizes traffic flow, reduces congestion, and improves mobility using real-time data. These case studies demonstrate successful examples of how integration challenges have been addressed.

\section{Case studies on dynamic rerouting implementation}

The case studies that demonstrate implementation of dynamic rerouting include the global navigation software Waze, which uses real-time user data to automatically reroute drivers based on traffic congestion \cite{laor2022waze}. Using user data, the app suggests alternative routes to save travel time and congestion, demonstrating dynamic rerouting in regular commutes. The London Congestion Charge Scheme (UK) dynamically reroutes vehicles by charging to control traffic flow in the city center during peak hours \cite{green2020london}. The scheme encourages drivers to use alternative routes, minimizing congestion and pollution. In addition, smart traffic control systems in Singapore dynamically redirect vehicles using real-time data from cameras, sensors, and GPS. Urban deployment of the technology reduces congestion by adjusting traffic signal timings and rerouting vehicles. Using dynamic rerouting, Curitiba's Bus Rapid Transit (BRT) system optimizes bus routes based on passenger demand \cite{borgman2020assessing}. Bus routes can be dynamically adjusted to meet customer needs and save journey time. Lastly, Japan's railway networks use dynamic rerouting algorithms to optimize train timetables and routes in reaction to natural catastrophes or system failures. Consequently, this results in minimal disruptions and efficient transit. In addition to these examples, a comprehensive study by \cite{aparicio2022assessing} on the robustness of multimodal transportation systems, exemplified through a case study in Lisbon, underscores the importance of dynamic rerouting in enhancing the resilience of urban transport networks. Their analysis provides valuable insights into how multimodal systems can adapt to changing circumstances while maintaining operational efficiency.

\begin{figure}
    \centering
    \includegraphics[width=15cm]{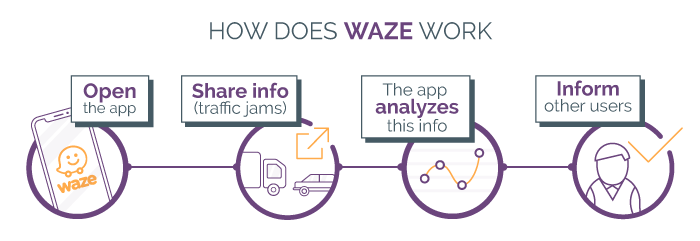} 
    \caption{Working of the Waze App. Adapted from \cite{s-2023}}
    \label{fig:fig4}
\end{figure}

Analyzing the case studies of successful implementation reveals distinctive results, lessons gained, and scalability of dynamic rerouting. The analysis outcomes reveal that most dynamic rerouting methods reduce congestion and travel time, improving transportation efficiency \cite{laor2022waze}\cite{green2020london}. Subsequently, efficient transportation systems boost production and lower operational expenses. The efficiency in transportation means that users have shorter commuter time and more convenience with other routes. Dynamic rerouting also enhances traffic flow optimization, reducing carbon emissions and pollutants in several solutions. Additionally, the case studies of implementation of successful implementation of provides numerous lessons to be learned from the implementation. Dynamic rerouting models promote user engagement. For instance, dynamic rerouting systems like Waze depend on user engagement \cite{laor2022waze}. An important lesson learned from the case study is that government laws and regulations support systems like London's Congestion Charge, making the dynamic rerouting more successful \cite{green2020london}. In addition, dynamic rerouting requires adaptability to real-time data and conditions. Lastly, dynamic rerouting solutions need infrastructural and technology investments to scale to ensure the success of the project. The scalability aspect derived from the case study is dynamic rerouting can scale globally; for instance, digital systems like Waze demonstrate global scalability \cite{laor2022waze}. Also, to note, dynamic rerouting methods scale differently depending on local infrastructure, regulatory climate, and user behavior. Lastly, AI, IoT, and big data technologies improve precision and real-time decision-making, enabling these solutions to scale.

\section{Impact of dynamic rerouting on sustainability and urban development}

Dynamic rerouting improves urban transport sustainability in diverse ways. Dynamic rerouting systems reduce idle and traffic delays, minimizing carbon emissions and pollutants. Improving air quality and reducing urban carbon footprints support environmental sustainability goals. Effective rerouting reduces fuel use and waste, promoting efficient use of resources. Efficiency in energy use also reduces environmental impact \cite{lv2023impacts}. In addition, dynamic rerouting commonly suggests public transit, cycling, and walking, promoting alternative modes of transport. Encouragement of these modes minimizes private vehicle use, reducing traffic and environmental damage. Urban planning and development is crucial in dynamic rerouting as the systems direct traffic away from residential and environmentally protected areas. Such planning supports urban growth while preserving the environment. Further, dynamic rerouting systems support green programs and ensure city-wide sustainability by connecting transportation practices with environmental aims. Lastly, dynamic rerouting systems reveal traffic patterns, congestion hotspots, and transit needs, which aid in data-driven decision-making. These findings inform urban planning, infrastructure, and public transit improvements, promoting sustainable growth.

Dynamic rerouting significantly impacts urban planning and development in diverse ways. Dynamic rerouting requires flexible infrastructure development, hence influencing infrastructure planning. Consequently, cities adopt smart signaling systems, adaptive traffic management infrastructures, and technological advances in roadways. Urban planners use rerouting data to determine zoning and land use. Therefore, rerouting traffic may reduce the effect in residential, green, and business areas. Dynamic rerouting affects transport-oriented development (TOD), which centers urban development along transportation hubs \cite{yang2019integrated}. TOD encourages walkability, public transit, and car reduction. In addition, dynamic rerouting suggests ideal routes and provides real-time updates to encourage public transit use, promoting public transit improvements \cite{zhao2018dynamic}. This data helps urban planners build public transit networks, enhancing accessibility and connectivity. Further, dynamic rerouting helps prepare evacuation routes or manage traffic during emergencies in vulnerable cities, facilitating emergency planning. Dynamic rerouting promotes urban technology integration since data-driven decision-making makes better municipal management. Lastly, dynamic rerouting shows that urban design must be technology-driven to accommodate expanding urban populations and handle environmental, social, and economic issues.
Different studies support the numerous benefits of dynamic rerouting. For instance, traffic management and intelligent transport systems found that dynamic rerouting can cut urban traffic congestion by significant margins. Dynamic rerouting also reduces CO2 emissions by diverting vehicles from high-pollution zones or providing eco-friendly routes \cite{lv2023impacts}. In addition, dynamic rerouting can save cities millions in traffic congestion and delay-related costs \cite{zhao2018dynamic}. A correlation exists between dynamic rerouting implementation and urban development. Cities with advanced rerouting systems invest more in mixed-use developments and public transit.

\section{Future possibilities and existing research gaps in dynamic rerouting}

Dynamic rerouting has great potential for adoption. The future possibilities of dynamic rerouting include AI and Machine Learning. Notably, future dynamic rerouting systems may use AI to predict traffic trends and learn from past data in real time. Sensor technology and data fusion techniques may increase data integration, providing more complete and accurate information for rerouting decisions. In addition, the future potential of dynamic rerouting is that the system may optimize routes for time efficiency, environmental effect, fuel consumption, and user experience. Future research aims to build rerouting algorithms that can quickly adjust to unexpected occurrences like accidents, weather changes, and infrastructure breakdowns. Lastly, dynamic rerouting will focus on user-centric rerouting, such as personalizing routing suggestions, incorporating user feedback, and offering personalized routing solutions that improve user satisfaction. 

A significant advancement in this field is the potential integration of large language models with dynamic rerouting of multimodal transport systems. The use of multimodal features and intelligent transportation systems (ITS) could be enhanced by the scalability and adaptability of multimodal composition frameworks. For example, integrating OpenAI's multimodal model with audio language models presents opportunities for the integration of large language models with multimodal transport systems \cite{liu-2023}. Such integration can benefit from advancements in multimodal features, intelligent transportation systems, and the scalability of multimodal composition frameworks. These approaches can enhance natural language understanding, optimize multimodal journeys, and improve participant management within multimodal transport systems.

Several research gaps emerge from the multimodal transportation network dynamic rerouting systematic review paper. Future research should consider ethics and social implications by focusing on data privacy, equity, and decision-making justice while rerouting algorithms. The inquest should investigate how these algorithms affect socioeconomic groups and the ethical implications of rerouting decisions. In addition, future research should build user trust by assessing dynamic rerouting proposals and approaches' reliability. The analysis should investigate how the systems make consistent, trustworthy rerouting judgments, especially in emergencies. Future research should focus on user-centric solutions by assessing dynamic rerouting system user approval, preferences, and satisfaction. The investigations should determine users' preferences and how systems can personalize rerouting proposals to promote happiness and acceptance. Further, a study on autonomous vehicle rerouting algorithm demands and problems should be conducted. The research should examine how autonomous car algorithms differ from normal transport modes and their unique features. Future investigations should inquire into how systems respond to abrupt and unpredictable occurrences in real-time to reroute efficiently without added disruptions. Lastly, future research should investigate ways to make rerouting systems more resilient to cyberattacks and maintain functionality. 

In conclusion, dynamic rerouting in multimodal transportation networks must be studied and researched. For a more efficient and sustainable future, gaps should be investigated to solve the gaps and advance the field. Future research on ethics, interdisciplinary collaboration security, continuous innovation, and policy and standardization should be explored. Finally, the analysis and recommendations of the review paper practitioners may improve dynamic rerouting systems and create a more efficient, sustainable, and user-centric transportation landscape.

\clearpage
\bibliographystyle{unsrt}  
\bibliography{references} 
\end{document}